# A Novel Hybrid Algorithm for Task Graph Scheduling

**Vahid Majid Nezhad[1], Habib Motee Gader[2] and Evgueni Efimov[3]**

**[1] Department of Computer Engineering, Islamic Azad University, Shabestar Branch
Shabestar, East-Azerbaijan, Iran**

**[2] Department of Computer Engineering, Islamic Azad University, Shabestar Branch
Shabestar, East-Azerbaijan, Iran**

**[3] United Institute of Informatics Problems of National Academy of Science of Belarus
Minsk, Belarus**

**Abstract**

One of the important problems in multiprocessor systems is Task Graph Scheduling. Task Graph Scheduling is an NP-Hard problem. Both learning automata and genetic algorithms are search tools which are used for solving many NP-Hard problems. In this paper a new hybrid method based on Genetic Algorithm and Learning Automata is proposed. The proposed algorithm begins with an initial population of randomly generated chromosomes and after some stages, each chromosome maps to an automaton. Experimental results show that superiority of the proposed algorithm over the current approaches.

## 1. Introduction

Although computer performance has evolved exponentially in the past, there have always been applications that demand more processing power than a single state-of-the-art processor can provide. To respond to this demand, multiple processing units are employed conjointly to collaborate on the execution of one application. Computer systems that consist of multiple processing units are referred to as parallel systems. In designing parallel systems different aspects have to be taken into consideration such as the manner of dividing a program into some tasks and the manner of tasks assignment to processors which is called Task Graph Scheduling.

Task Graph Scheduling is an important issue in the distribution of programs on the processors of a parallel system. Because task graph scheduling is an NP-Hard problem, methods of random search are utilized for finding the nearly optimal scheduling [1]. Among the various methods of random search, Genetic Algorithm (GA) has been one of the best ones ever used for Task Graph Scheduling [2-6]. Learning Automata (LA) is another method that is used for Task Graph Scheduling [7-9]. Also other methods are also used for Task Graph Scheduling that we are going to consider some of them in this paper [10-12].

In this paper parallel programs are presented by the Task Graph. Fig. 1 depicts an example of the task graph for a program. The numbers allocated to the graph nodes represent the costs of the completion of that node, and the numbers given to the manes of the graph represent the connection cost among nodes. Each contrastive node is a task.

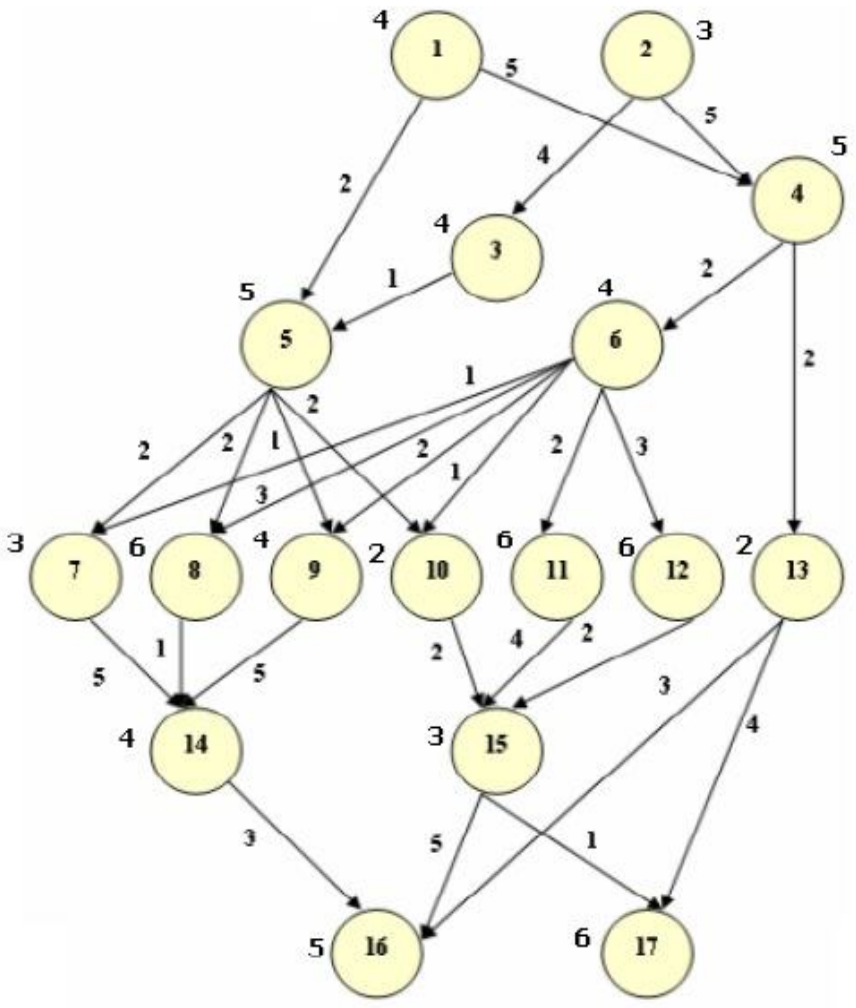

Fig. 1 Example of task graph with 17 tasks.

The connection cost between two nodes is put forward when the instruction of those two nodes are applied in different processors. If both instructions are running on same processor, in that case, the connection cost between those two instructions is considered zero (in fact reality the connection cost is not zero but very minimal time. Due to the meagerness of this cost, it is considered zero). The rest of the paper will be as follows:

In part two, an outline of the Genetic Algorithm is put forward, afterwards, in part three learning automata is introduced. After that, in part four, the new algorithm for solving the problem of the task graph scheduling is presented and in part five, result of experiments are analyzed and then, in final part conclusions will be investigated.

## 2. GENETIC ALGORITHM

Genetic Algorithms which act on the basis of evaluation in nature search for the final solution among a population of potential solution. In every generation the fittest of that generation selected and after reproduction produce a new set of children. In this process the fittest individuals will survive more probably to the next generations. At the beginning of algorithm a number of individuals (initial population) are created randomly and the fitness function is evaluated for all of them. If we do not reach to the optimal answer, the next generation is produced with selection of parents based on their fitness and the children mutates with a fixed probability then the new children fitness is calculated and new population is formed by substitution of children with parents and this process is repeated until the conclusion condition is established.

The most advantages of this algorithm compared with common methods are: parallel search instead of serial search, not requiring any additional information such as problem solving method, in-deterministic of algorithm, easy implementation and reaching to several choices. GA uses several operators, each of which have different types and can be implemented using different methods.

## 3. LEARNING AUTOMATA

Learning in LA is choosing an optimal action from a series of allowable automata actions. This action is applied on a random environment and the environment gives a random answer to this action of automata from a series of allowable answers. The environment's answer depends statistically on automata action. The term environment includes a set of outside conditions and their effect on automata operation. Connection of an automaton with the environment is shown in Fig. 2. In this paper the used automata is an Object Migration Automata (OMA).

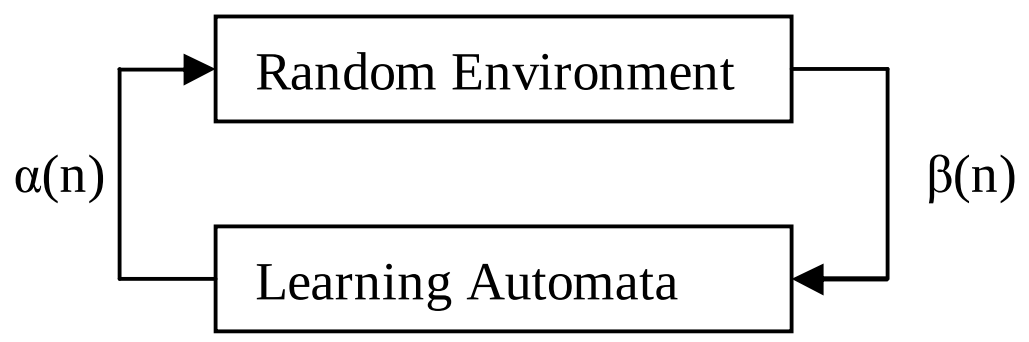

Fig. 2 Connection of LA with random environment.

## 4. The Proposed algorithm

In the proposed algorithm, the combination of genetic algorithm (GA) and learning automata (LA) are used. One of the most important features of the GA is that it has stochastic behavior which is because of the genes too much change. Therefore it is unstable but it has a high speed in creating an approximately appropriate population of chromosomes. The significant feature of LA is its stability because actions or genes don’t change too much. In other words, in each stage of LA rewarding and penalizing are done. While rewarding, a gene strengthens and doesn’t replace. While penalizing, a gene may get weakened or it may be replaced.

The base of proposed algorithm is that, in the first stages, GA be used. By too much change in genes we can reach to an approximately appropriate population of chromosomes. After that to avoid instability and stochastic behavior of GA, the chromosomes are mapped to automata and in order to make it stable, other stages are done by LA. It means in our proposal algorithm the advantage of both method are used.

In details, proposed algorithm mixes GA and LA as follow: First for running genetic algorithm, some chromosomes as initial population are produced. One of these chromosomes is displayed in Fig. 3. As shown in Fig. 3 the genes from left to right indicates first task, second task, … and ninth task. And a random number is assigned to each gene so that the random numbers indicate two concepts:

- The priority of tasks. Greater numbers have more priority.
- The number of the processor that is in charge of running that task. For specifying the number of the processor, we must use the mod of random number to total number of processors.

Second, for running Learning automata each chromosome maps to an automaton. For this propose each gene of chromosome convert to an action of automaton. For example the chromosome of Fig. 3 is converted to an automaton in Fig. 4.

| T1 | T2 | T3 | T4 | T5 | T6 | T7 | T8 | T9 |
|---|---|---|---|---|---|---|---|---|
| 15 | 11 | 14 | 9 | 17 | 5 | 3 | 2 | 7 |

Fig. 3 An instance of chromosomes.

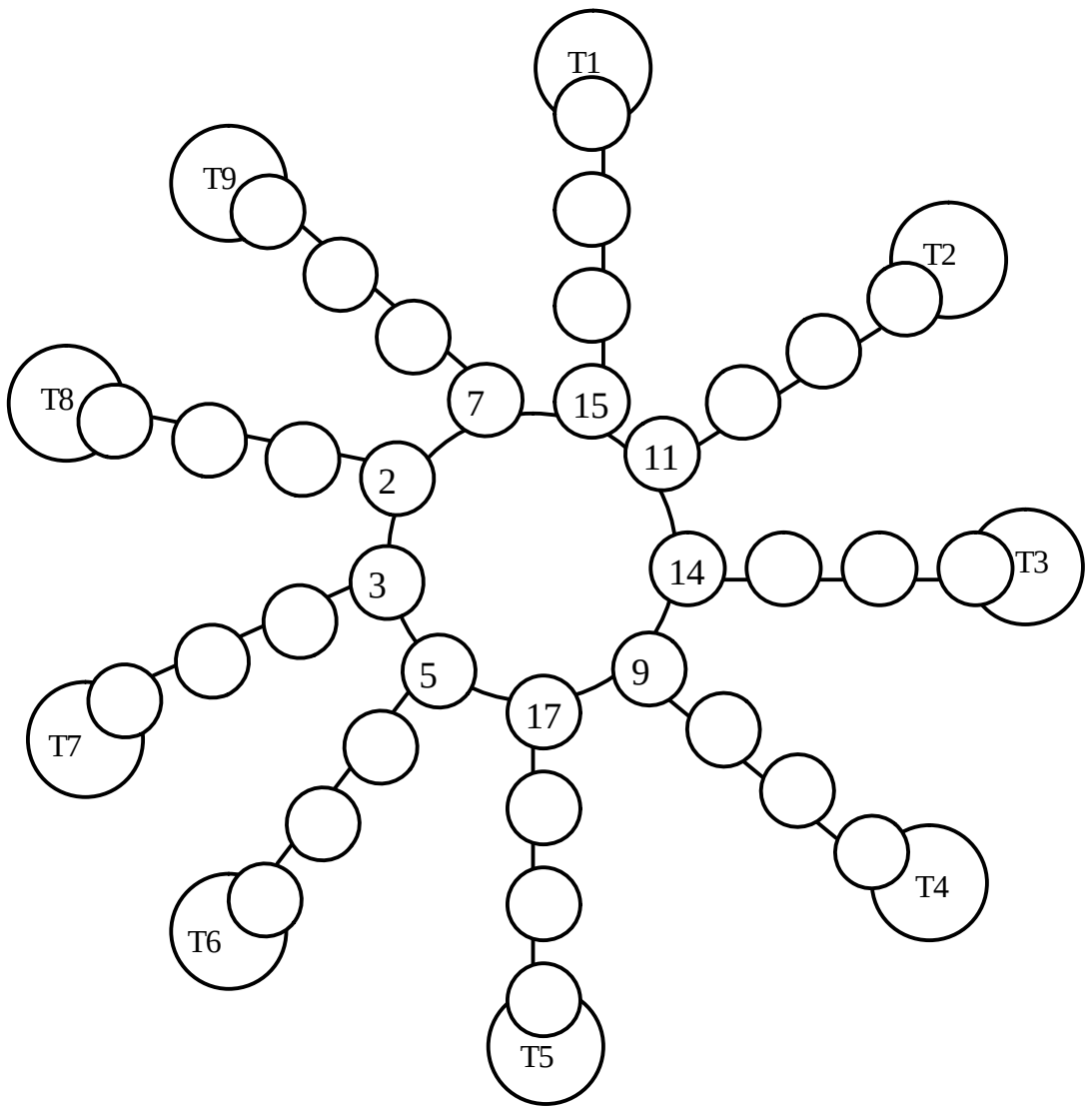

Fig. 4 Mapping a chromosome to an automaton.

Next, details of genetic algorithm about fitness function, crossover operator, mutation operator and selection operator in my proposed algorithm are described.

In Genetic Algorithm, fitness function determines whether chromosomes are going to stay alive or not. In the problem of task scheduling, the object is to find a short makespan. Eq. (1) Shows Fitness function for evaluation of chromosomes.

$$f_k = {}^{1}\!/\!_{m_k}, k = 1,2,....., \; popsize \qquad (1)$$

$m_k$ : the makespan resulting from *k th* chromosome.
*popsize*: population size.

In this article, a novel method for crossover operator has been described. The combination method used in this article is a two-point one. First two points are randomly chosen as subclasses, and then their contents and orders are analyzed. For instance, as shown in Fig. 5 the substring chosen from first chromosome, has a weight order of 1-2-3-4. This weight order is used for changing the substring chosen by second chromosome. Thus, the 6-13-15-11 is changed to 15-13-11-6. WMX algorithm is not one, which changes only the contents of two points selected from two chromosomes, but it also changes the contents of classes according to weight priorities.

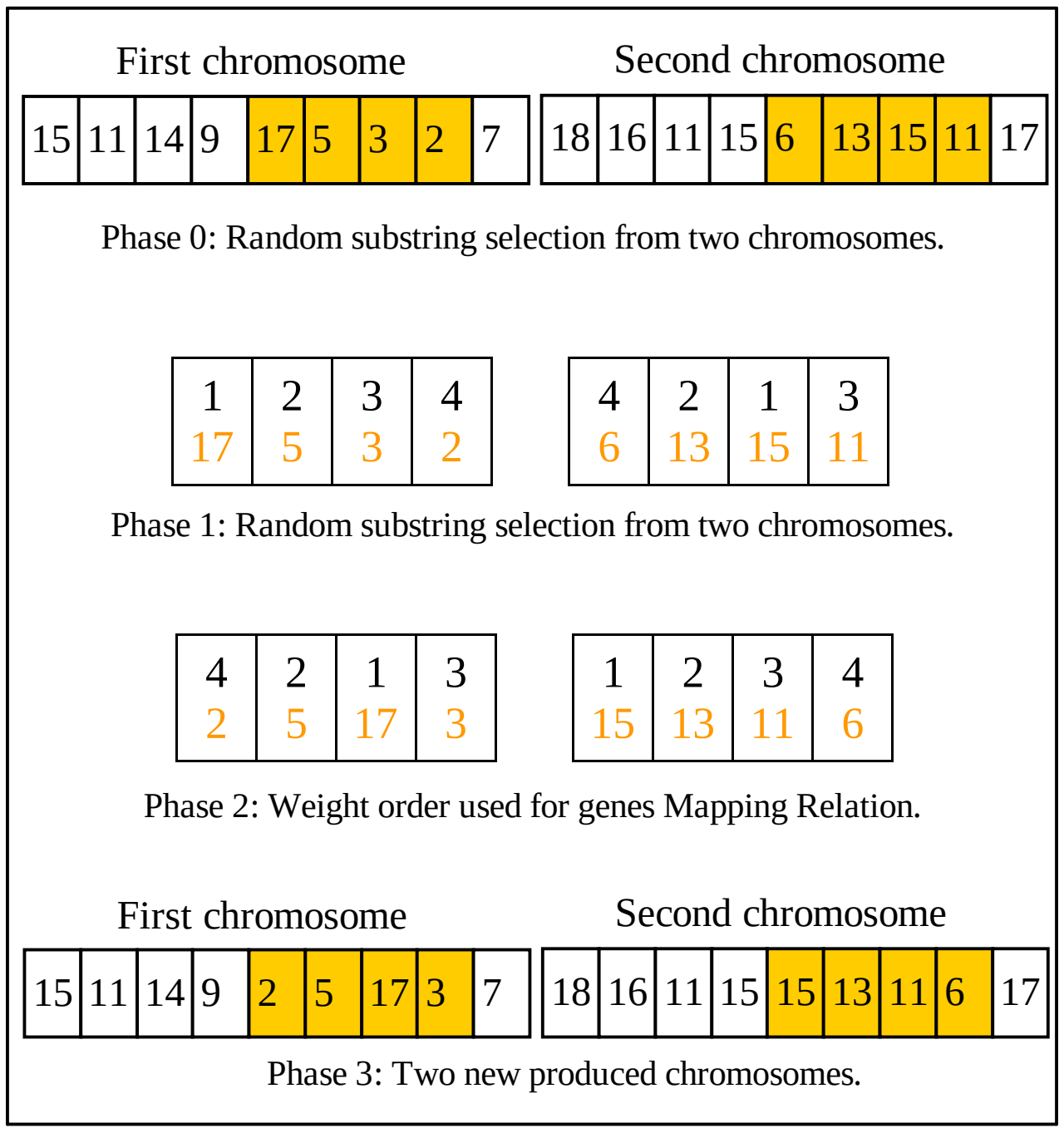

Fig. 5 Example of Crossover Operator.

For operating mutation, two genes are randomly selected from a chromosome and their amounts are changed with each other.

Selection operator in this article is as follows: In each step of new population production, a percent of chromosomes, which has least amount of fitness, are selected and enter in the new population directly. The rest of the population is produced through combining chromosomes.

Next details of automata and its operators are described. In this automaton $\alpha$ $\{\alpha_1,...,\alpha_k\}$ is the set of allowed action for the learning automata. This automaton has $k$ actions (i.e. the number of the actions of this automaton is equal to number of the tasks of the graph). Each action specifies a special task when and where will be executed. $\phi\{\phi_1, \phi_2, \phi_3 ..., \phi_{KN}\}$ is the set of situations, and $N$ is the memory depth for automata. The situation set of this automaton is divided to $k$ subsets and each task is categorizing to where and which position it is located. In the set of $j$'s action, position $\phi_{(j-1)N+1}$ is called internal position and $\phi_{jN}$ position is called boundary position. A nodes in $\phi_{(j-1)N+1}$ position is called is called a more important node, and a node in $\phi_{jN}$ position is called a less important node.

Since, each chromosome is presented as a learning automaton, in each automaton, after considering the fitness of a gene (either processor or action), which is selected on a random basis, that gene is duly penalized or rewarded. As a result of penalizing a gene, its position in the boundary position of an action, leads to a change in its action and, in consequence, creation of a new makespan. Reward action occurs when the fitness of a task is smaller than its threshold. Eq. (2) shows fitness of $t_i$ and Eq. (3) shows threshold rate of $t_i$.

$$f(t_i) = x_i / y_i \qquad (2)$$

$$Th(t_i) = r_i / N \qquad (3)$$

Eq. (4) And Eq. (5) show $x_i$ and $y_i$ equations. $x_i$ is the sum of connection cost of all parent and offspring nodes of $t_i$ node so that $p_{ti} \neq p_{tj}$ and $y_i$ is the sum of the connection costs of all parent and offspring nodes of $t_i$ node.

$p_{ti}$ : A processor that $t_i$ task is performed on it.

$p_{tj}$ : A processor that $t_j$ task is performed on it.

$c(t_i, t_j)$ : Communication cost between $t_i$ and $t_j$ tasks.

$N$: The number of all graph tasks.

$r_i$ : Consist of a number of related tasks to $t_i$ task that is executed on a processor which $t_i$ task is run in it.

$$x_i = \sum c(t_i, t_j) \quad if \quad p_{ti} \neq p_{tj} \qquad (4)$$

$$y_i = \sum c(t_i, t_j) \qquad (5)$$

$r_i$ has a reverse relation with $x_i$; as $r_i$ increases $x_i$ decreases and vice versa. If the fitness level of a $t_i$ task is equal to zero, it means that all related tasks of $t_i$ are performed on the same processor. Therefore, the lower value of fitness is better for scheduling problem. In case the fitness level of a task is more than the threshold amount, then the task gets penalized. Two positions are possible when penalizing a task:

*a)* The task's value might be in a position other than boundary position. In this case, penalizing makes it less important. How the task's value of $t_3$ task is penalizes, is shown in Fig. 6.

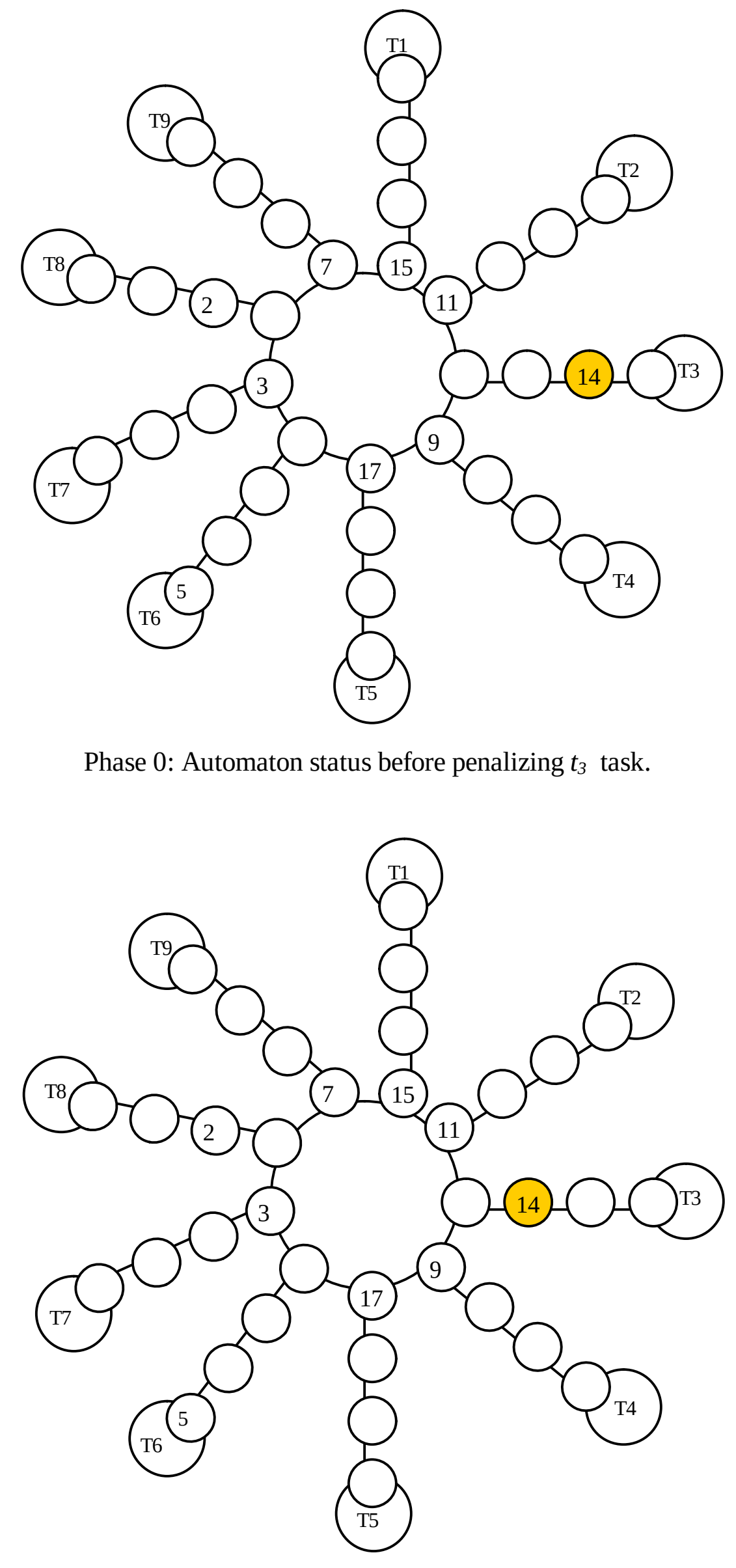

Phase 0: Automaton status before penalizing $t_3$ task.

Phase 1: Automata status after penalizing $t_3$ task.

Fig. 6 $t_3$ task penalizing.

*b)* The task's value might be in boundary position. In this case, we look for a task in the graph that has the greatest reduction in the amount of fitness when the values of them are changed. Now if the value of found task is in the boundary position, two values are changed with each other and if otherwise, i.e. if the value of found task is not in the boundary position, first the value of found task should be moved to its boundary position and then values change occurs. Fig. 7 shows how $t_4$ task is penalized.

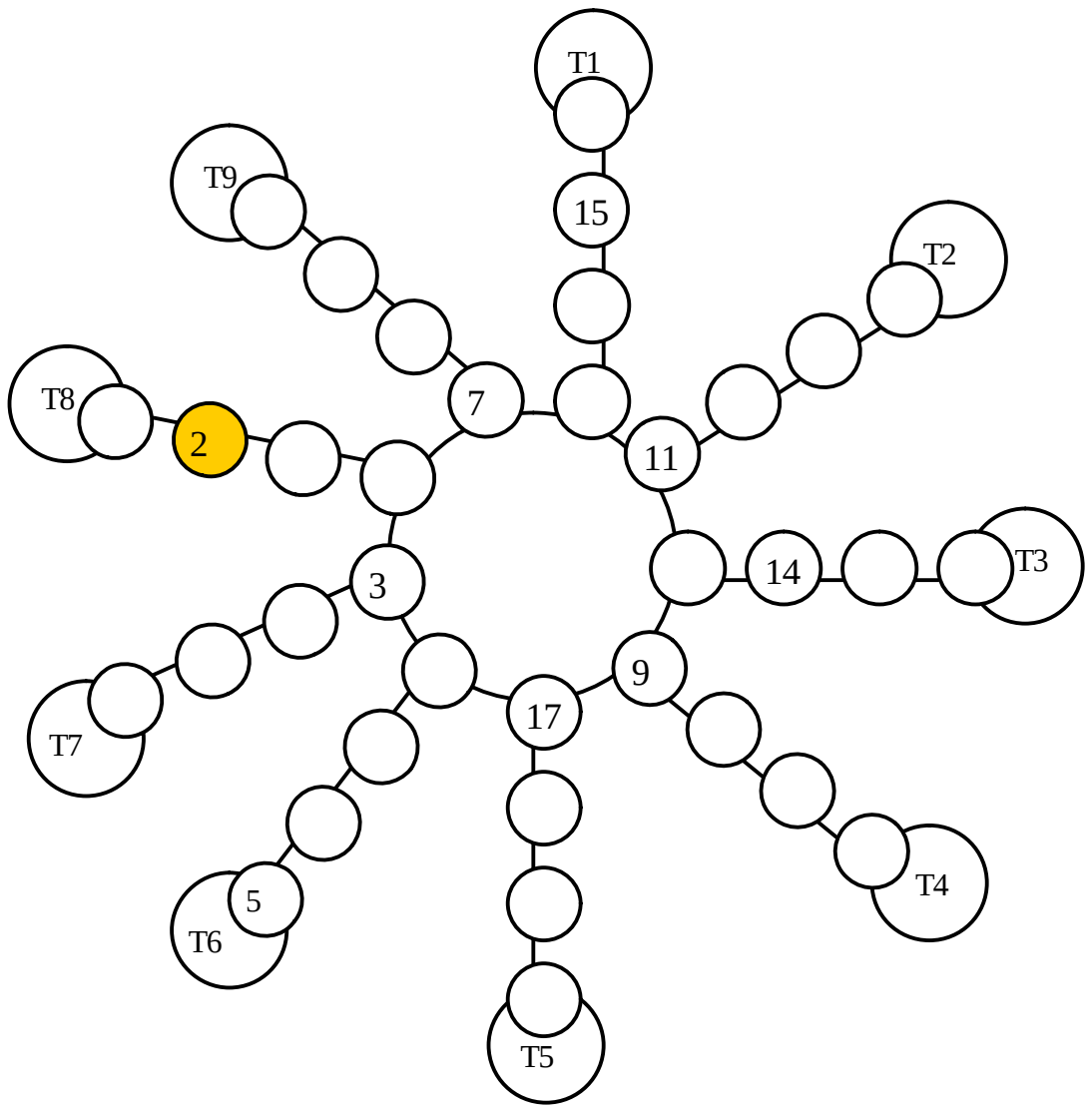

Phase 0: $t_4$ task status before penalizing.

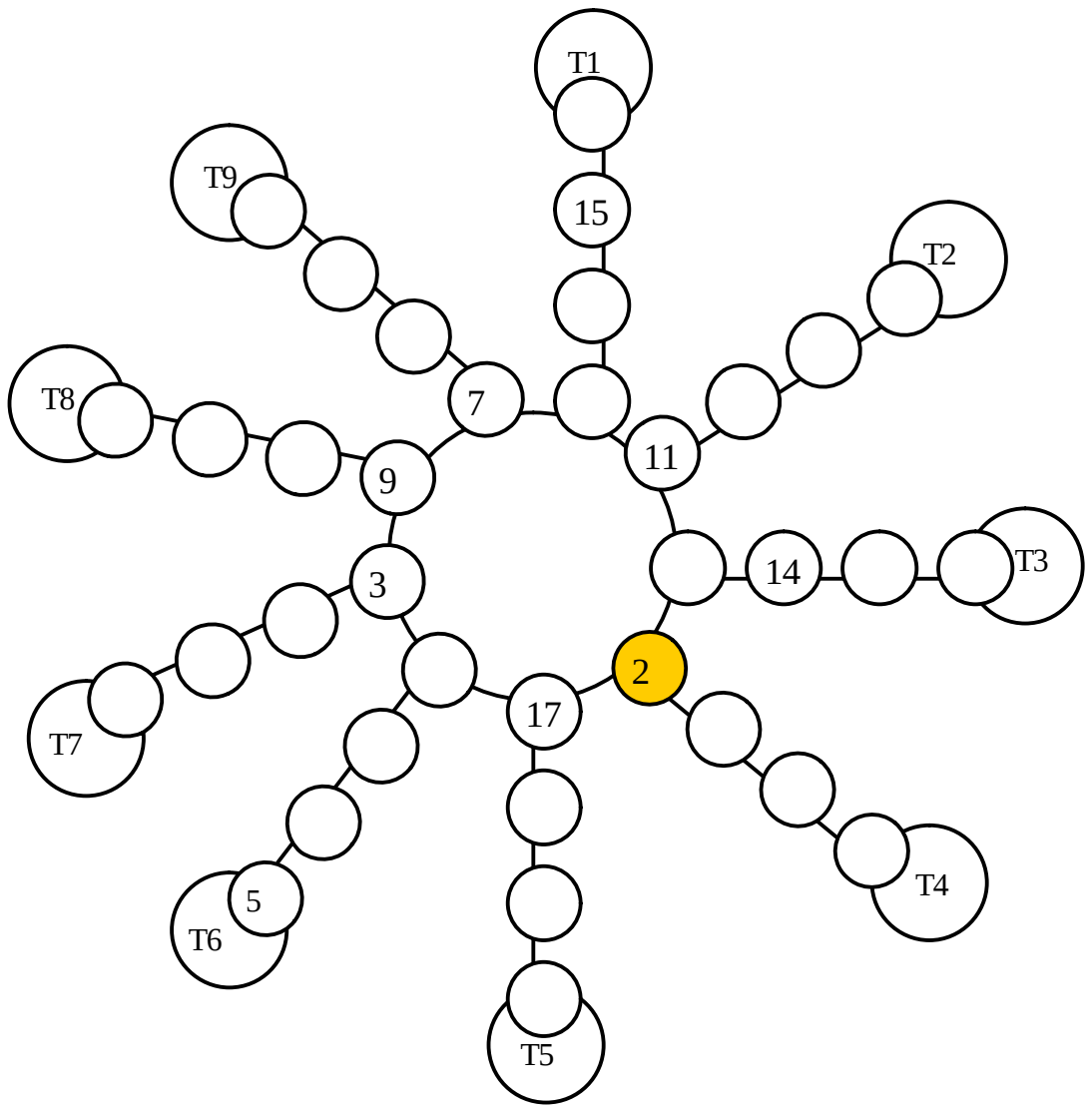

Phase 1: Value of $t_4$ transmit to boundary position and values of $t_4$ and $t_8$ tasks are changed.

Fig. 7 $t_4$ task penalizing.

# 5. SIMULATION RESULTS

In this article, the performance of the proposed algorithm is compared with well-known definite and indefinite algorithms. Parameters that are used in PMC_GA and the proposed algorithm are shown in table 1. Next, three experiments are described and simulation results are investigated.

Table 1: Algorithms Parameters

| Algorithm | Memory Depth | Mutation Rate | Crossover Rate | Population |
|---|---|---|---|---|
| PMC_GA | - | 0.3 | 0.7 | 100 |
| Proposed | 5 | 0.3 | 0.7 | 100 |

Test Algorithms which are used in this section are: MCP (modified critical path) by Wu and Gajski [10], DSC (dominant sequence clustering) by Yang and Gerasoulis [11], MD (mobility directed) by Wu and Gajski [10], DCP (dynamic critical path) by Kwong and Ishfaq [12], PMC_GA by Hwang, Gen and Katayama [13].

First experiment: by observing the task graph in Fig. 8, results obtained from various algorithms [13] and the proposed algorithm is displayed in table 2. Also acquired Gantt chart of the proposed algorithm is shown in Fig. 9. It becomes evident that the proposed algorithm reaches the better results in fewer generations.

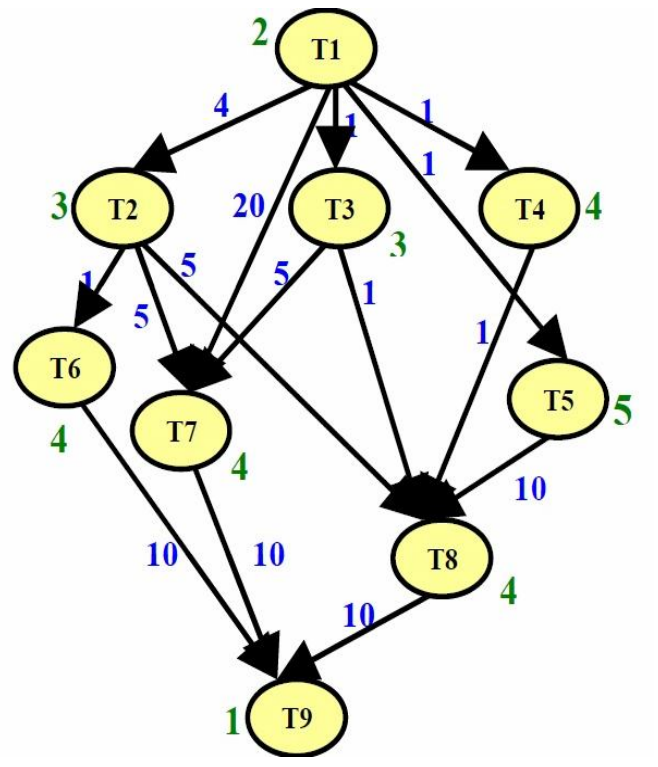

Fig. 8 Example of task graph with 9 tasks [13].

Table 2: Comparative results of the proposed algorithm with others.

| Algorithms | MCP | DSC | MD | DCP | PMC_GA | Proposed |
|---|---|---|---|---|---|---|
| No. Processors | 3 | 4 | 2 | 2 | 2 | 2 |
| Finish Time | 29 | 27 | 32 | 32 | 23 | 21 |
| Iterations | - | - | - | - | 50 | 15+25 |

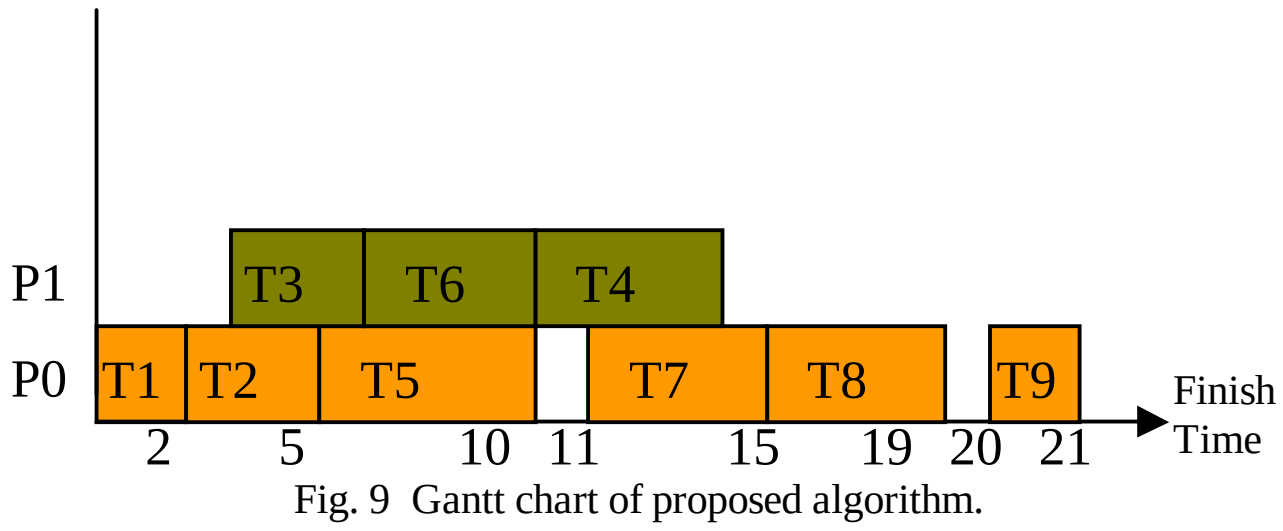

Fig. 9 Gantt chart of proposed algorithm.

Second Experiment: the second experiment is performed on the graph of Fig. 10 and the results obtained from various algorithms [13] and the proposed algorithm are shown in table 3. It can be seen that the proposed algorithm reaches the response in fewer generations.

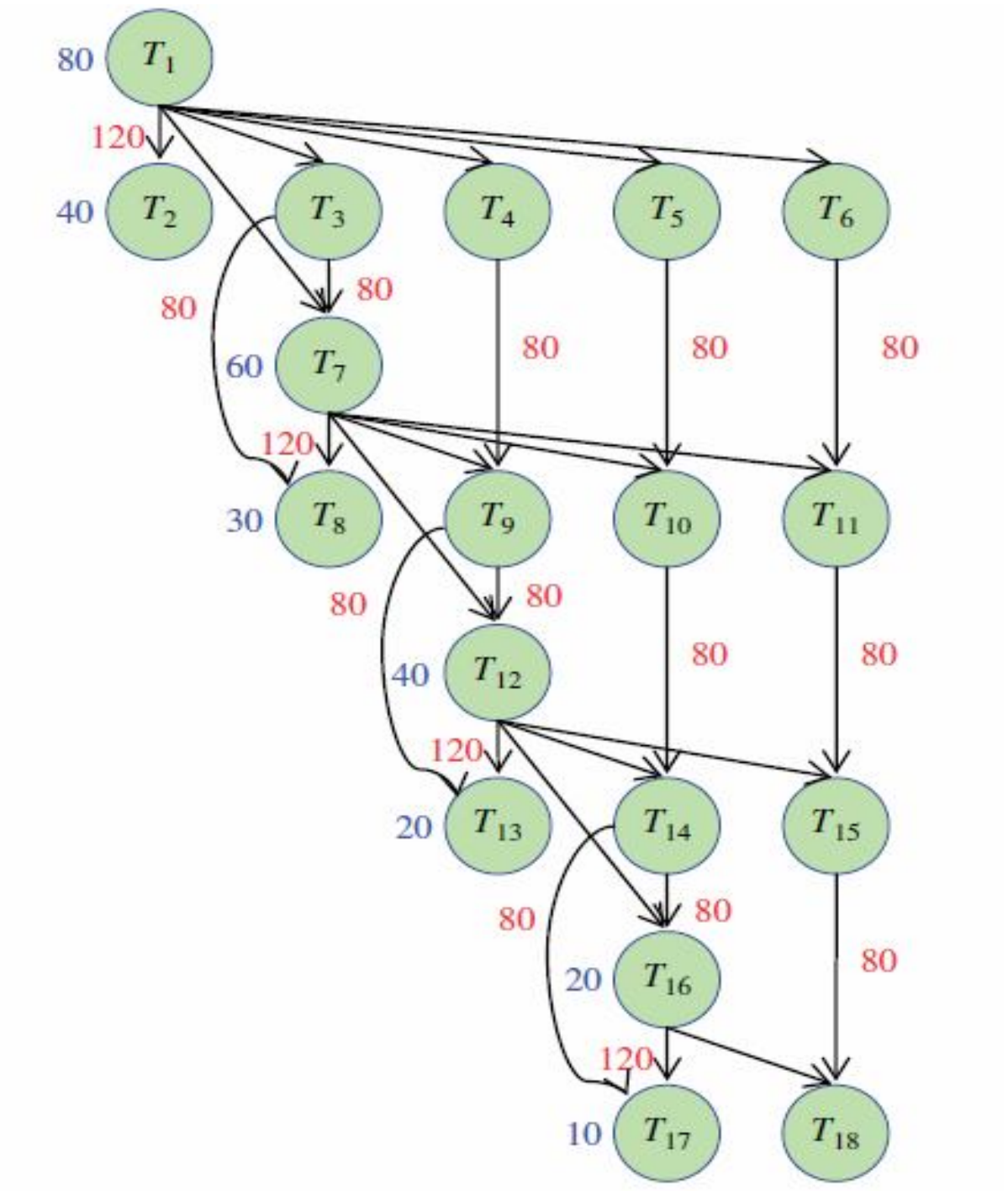

Fig. 10 Example of task graph with 18 tasks [13].

Table 3: Comparative results of the proposed algorithm with others.

| Algorithms | MCP | DSC | MD | DCP | PMC_GA | Proposed |
|---|---|---|---|---|---|---|
| No. Processors | 4 | 6 | 3 | 3 | 2 | 2 |
| Finish Time | 520 | 460 | 460 | 440 | 440 | 440 |
| Iterations | - | - | - | - | 100 | 30+40 |

Third Experiment: for testing the proposed algorithm and comparing it with the PMC_GA [13] on a larger DAG, the simulations are performed in different conditions and based on some standard task graph database [14]. Note that we add some communication cost to the database graphs and make some graph with communication cost to test our proposed in a real condition and compare it with the PMC_GA. Also terminating condition of both methods is 10 iterations with same fitness. The simulation results are shown in table 4. It becomes evident that the proposed algorithm in comparison with PMC_GA reaches better results.

Table 4: Comparative results with 50 tasks graphs (rnc50.tgz, rand0010stg, rand0016stg) [14]

| Algorithms | PMC_GA | Proposed |
|---|---|---|
| No. Processors | 2 | 2 |
| Finish Time for mc50.tgz,rand0010 | 133 | 115 |

## 6. Conclusions

In this paper the hybrid algorithm is proposed for Task Graph Scheduling in parallel systems. This algorithm utilizes advantages of Genetic Algorithm and Learning Automata methods to search into the state space. In proposed algorithm by using good initial population of PMC_GA and stability of Learning Automata in search process, the number of generations needed for reaching the optimal response decreases. Also the results of the experiments show that the proposed algorithm from optimal response point of view acts better than other methods. Therefore, the results of the experiment show the superiority of the proposed algorithm to current algorithms.

### Acknowledgments

This paper was under grant and supported by the Islamic Azad University, Shabestar Branch. The authors wish to express their thanks for the support.

**Vahid Majid Nezhad** received the B.Sc. degree in Software Engineering from the Islamic Azad University, Tabriz Branch, Iran, in 2004, and the M.Sc. degree in Software Engineering from the Islamic Azad University, Tehran Branch, Iran, in 2007. Since 2007, he has been with the Faculty of Computer Eng., at the Islamic Azad University, Shabestar Branch, Iran, where he is currently an Academic Staff. His research interests include Speech Processing and Optimization Problems.